\documentclass[12pt]{article}
\usepackage{graphicx}
\usepackage{latexsym}
\usepackage{amssymb}
\begin{document}
\hspace {-8mm} {\bf \LARGE Local-density approximation for exchange energy functional in 
excited-state density functional theory} \\ \\

\hspace {-8mm} {\bf Prasanjit Samal and Manoj K. Harbola}\\
{Department of Physics, Indian Institute of Technology,
 Kanpur 208016, India}
\\
\begin{abstract}
An exchange energy functional is proposed and tested for obtaining
a class of excited-state energies using density-functional formalism.
The functional is the excited-state counterpart of the local-density
approximation functional for the ground-state. It takes care of the
state-dependence of the energy functional and leads to highly accurate
excitation energies.
\end{abstract}

\newpage
\section{Introduction}
Success of density functional theory (DFT) \cite{gross1,parr1} for the 
ground-state calculations had prompted search 
\cite{ziegler,Gunnar,vonB,LP1,rkp,theo,olivi1,nagy} 
for similar theories for the excited-states. Over the past decade, time-dependent 
density-functional theory (TDDFT) \cite{gross2} has become a standard tool 
\cite{casida,peters,gauss} for obtaining transition energies and the associated 
oscillator strengths. However, despite it's widespread use, the theory is not without 
limitations. For example, if adiabatic approximations are applied for the 
exchange-correlation kernel, excitation energies for double excitation of electrons
cannot be obtained \cite{tozer} within TDDFT; obtaining these still remains \cite{burke}
a challenge in the TDDFT approach. At the same time, the charm of getting the 
excitation energy as the difference between two total energies remains. This is because
one can choose the excited-state at will, promoting as many electrons as one wishes to 
a set of chosen orbitals, calculate the corresponding total energy and find the 
excitation energy by subtracting the ground-state energy from it; we refer the reader 
to the works of Ziegler et al \cite{ziegler}, von Barth \cite{vonB} and Gunnarsson and 
Jones \cite{GunnarJ} for some of the early attempts to apply this approach to study low
excitation energies. Thus research in the direction of performing a Kohn-Sham like 
calculation for the excited-states continues.
 
A ground-state like DFT approach to obtain the total energy
of an excited-state has been developed by G\"{o}rling \cite{gorling}
and by Levy and Nagy \cite{levy1}. The theory is based on the 
constrained-search approach \cite{levy2} and proposes that the energy of an 
excited-state can also be written as a functional
\begin{equation}
E[\rho] = F[\rho,\rho_{0}] + \int\rho({\bf r})v_{ext}({\bf r})d{\bf r}
\label{1}
\end{equation}
of the excited-state density $\rho({\bf r})$. Here $F[\rho,\rho_{0}]$ is a
bi-density functional that depends on the ground-state density $\rho_{0}$
also, and $v_{ext}({\bf r})$ is the external potential that the electrons
are moving in. The bi-density functional for the density $\rho$ of the $nth$
excited-state is defined via the constrained-search formulation as
\begin{equation}
F[\rho,\rho_{0}] = {\rm min}_{\Psi \to \rho} \left< \Psi | {\hat T} + 
{\hat V}_{ee}|\Psi \right> \;,
\label{2}
\end{equation}
where $\Psi$ is orthogonal to the lower $(n-1)$ states of the Hamiltonian, already
determined by the density $\rho_{0}$. Such a way of obtaining the functional
$F[\rho,\rho_{0}]$ makes it non-universal and also state-dependent. The
exchange-correlation energy functional $E_{xc}[\rho,\rho_{0}]$ for the
excited-state is then obtained by subtracting from $F[\rho,\rho_{0}]$
the non-interacting kinetic energy $T_{s}[\rho,\rho_{0}]$ and  the 
Coulomb energy $U_{Coul}[\rho]$ corresponding to $\rho$. 
The non-interacting kinetic energy $T_{s}[\rho,\rho_{0}]$ is defined in a manner 
similar to Eq.~(\ref{2}) by dropping the operator
${\hat V}_{ee}$ from the right hand side.  Thus (for brevity, from here onwards
we drop $\rho_{0}$ from the argument of the functional)
\begin{equation}
E_{xc}[\rho] = F[\rho]-T_{s}[\rho]-U_{Coul}[\rho]\;
\label{3}
\end{equation}
With the assumption that the excited-state density is non-interacting v-representable,  
the density is obtained by solving the excited-state Kohn-Sham equation (atomic
units are used throughout the paper)
\begin{equation}
\left[-\frac{1}{2}\nabla^{2}+v_{ext}({\bf r}) + \int\frac{\rho({\bf r'})}
{|{\bf r}- {\bf r'}|}d{\bf r'}+v_{xc}({\bf r})\right]\phi_{i}({\bf r})=
\epsilon_{i}\phi_{i}
({\bf r})
\label{4}
\end{equation}
as
\begin{equation}
\rho({\bf r}) = \Sigma_{i} {\rm n}_{i}|\phi_{i}({\bf r})|^{2}\;,
\label{5}
\end{equation}
where ${\rm n}_{i}$ is the occupation number of the orbital $\phi_{i}$.
In Eq.~(\ref{4}) the various terms have their standard meaning with $v_{xc}({\bf r})$ 
representing the exchange-correlation potential for the excited-state. It is 
determined by taking the functional derivative of the excited-state 
exchange-correlation energy functional. That a Kohn-Sham like calculation can be 
performed for the excited-states was first proposed by Harbola and Sahni \cite{hs} 
on physical grounds, and has been put \cite{ssms} on a rigorous mathematical footing 
recently on the basis of differential virial theorem \cite{holas}. Calculations of 
excited-state energies based on the Harbola-Sahni work have yielded excellent results 
\cite{sen,deb}. The near exact exchange-correlation potential for the singlet 
$1s2s\;^{1}S$ and triplet $1s2s\;^{3}S$ excited-state of Helium has also been 
constructed \cite{harb1,lind} recently. 
However, we are not aware of any work where an exchange-correlation functional for the 
excited-states has been reported; In performing excited-state calculations 
\cite{gorling,levy1,harb2}, either the ground-state functionals or the orbital 
based-theories \cite{hs,opm} have been employed. The proposition for the construction
of an excited-state exchange-correlation functional is indeed a difficult one 
since the functional is non-universal and also state-dependent.  Thus a general
functional form for it may not exist.

Against such a background, we ask if it is at all possible to obtain a simple
LDA-like functional for the excited-states.  To keep matters simple, we have
been looking at this problem within the exchange-only approximation. In this 
paper we show that it is indeed possible to construct an exchange
energy functional that gives transition energies comparable to the exact
exchange-only theories such as Hartree-Fock \cite{fischer}, optimized potential
\cite{opm} or the Harbola-Sahni \cite{sahnib} theory. The construction of the 
functional is based on the homogeneous electron-gas and in finding the final form 
of the functional we are guided mostly by qualitative plausibility arguments. Our 
work is thus exploratory in nature and represents probably the first attempt to 
construct an excited-state exchange-energy functional in terms of the density. 
The evidence of the accuracy of the functional constructed by us is given by the 
results of the transition energies of a large number of excited-states.  We also 
refer the reader to ref.~\cite{gorling2} for an expression for the change in the 
exchange energy in terms of the ground-state Kohn-Sham orbitals when an electron
is promoted from a lower energy orbital to a higher one.  

In the present work we take a particular class of excited states in which some
core orbital are filled, then there are some vacant orbitals and again there are 
some filled orbitals. We construct an LDA-like functional for such states in the
following section.

\section{Construction of the functional}

\begin{figure}[thb]
\includegraphics[width=5.5in,angle=0.0]{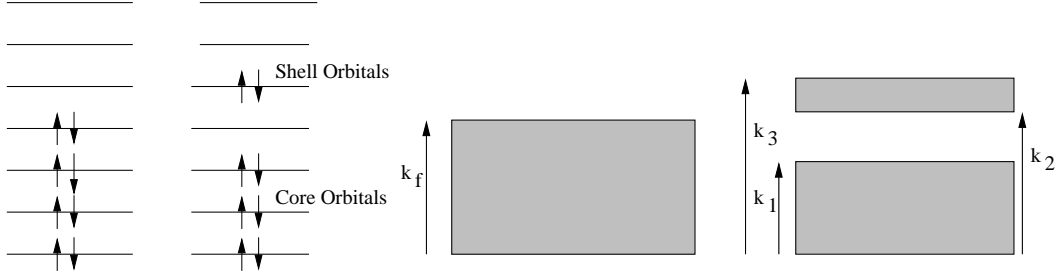}
\caption{Orbital and the corresponding $k-space$ occupation in the ground and the 
excited state configuration of a homogeneous electron gas(HEG).}
\label{k-space}
\end{figure}

As stated above, we now consider such excited-states where the occupation of the 
orbitals is such that the electrons occupy some core orbitals and some shell orbitals, 
leaving the orbitals between the core and the shell regions vacant.  This is shown 
schematically in Fig.~(\ref{k-space}). 
Such an excited-state would be obtained, for 
example, if an electron from the filled orbitals of the ground-state is excited to just 
above the occupied levels. The exact exchange energy for a set of occupied orbitals 
is given as
\begin{equation}
E_{X} = -\frac{1}{2}\sum_{\sigma} \sum_{i}^{occ} \sum_j^{occ} \left<\phi_{i\sigma}
{({\bf r}_{1})}\phi_{j\sigma}{({\bf r}_{2})} \left|\frac{1}{ |{\bf r}_{1}-{\bf r}_{2}|} 
\right|\phi_{j\sigma}{({\bf r}_{1})}\phi_{i\sigma}{({\bf r}_{2})} \right>\;.
\label{6}
\end{equation}
So the excited-state exchange energy when an electron is transferred
from one of the orbitals occupied in the ground-state to the lowest unoccupied
level is 
\begin{eqnarray}
E_{X}^{excited} & = & E_{X}^{ground} + {\sum_{j(\sigma_j=\sigma_{rem})}}
\left<\phi_{rem}{({\bf r}_{1})}
     \phi_{j}{({\bf r}_{2}) } \left|\frac{1}{ |{\bf r}_{1}-{\bf r}_{2}|}\right|\phi_{j}
     {({\bf r}_{1})}\phi_{rem}{({\bf r}_{2})} \right> \nonumber \\
             &   & -\frac{1}{2}\int\int\frac{|\phi_{rem}({\bf r}_{1})|^{2}|
\phi_{rem}
({\bf r}_{2})|^{2}}{|{\bf r}_{1}-{\bf r}_{2}|}d{\bf r}_{1}d{\bf r}_{2} \nonumber \\
        & & -\frac{1}{2}\int\int\frac{|\phi_{add}({\bf r}_{1})|^{2}|\phi_{add}
({\bf r}_{2})|^{2}}{|{\bf r}_{1}-{\bf r}_{2}|}d{\bf r}_{1}d{\bf r}_{2} \nonumber \\
                &   & - {\sum_{j(j\neq add)(\sigma_j=\sigma_{add})}} 
\left<\phi_{add}{({\bf r}_{1})}
     \phi_{j}{({\bf r}_{2}) } \left|\frac{1}{ |{\bf r}_{1}-{\bf r}_{2}|} \right|\phi_{j}
     {({\bf r}_{1})}\phi_{add}{({\bf r}_{2})} \right>\;, 
\label{7}
\end{eqnarray}
where $\phi_{rem}$ represents the orbital from which the electron has been removed 
and $\phi_{add}$ where the electron is added. The sum over the index $j$ in the 
second term on the right hand side runs over all the orbitals that are occupied
in the ground-state and $\phi_{add}$. On the other hand the sum in the fifth term 
runs only over the orbitals occupied in the ground-state.
We now attempt to make an LDA-like approximation for the excited-state exchange energy 
so that the difference (the last four terms in the equation above) between the 
approximate excited- and ground-state exchange energies is close to that given by the 
exact expression above. In making this approximation accurate, it is evident that the
self-energy terms (third and fourth terms on the right hand side of Eq.~(\ref{7}))
for the orbitals $\phi_{rem}$ and $\phi_{add}$ are to be treated accurately.

As the first step towards an excited-state functional, we make the correspondence
between the excited-states that we are considering and similar excitations in a
homogeneous electron gas (HEG). If the HEG is in it's ground state, the electrons are  
filled up to the Fermi level so that the electrons occupy wave-vectors in $k-space$ from 
$k=0$ to $k_{f}=(3\pi^{2}\rho)^{\frac{1}{3}}$, where $\rho$ is the electron density.
On the other hand, in an excited state of the system the electrons will occupy $k-space$ 
differently from the ground state. For the kind of excited-states that we consider in 
this paper, the corresponding occupation in the $k-space$ is as follows:
The electrons occupy orbitals from $k=0$ to $k_{1}$ and $k_{2}$ to $k_{3}$ 
with a gap in between as shown in Fig.~(\ref{k-space}).
So the excited state density is given by
\begin{equation}
\rho = \rho_{c}+\rho_{s}
\label{a}
\end{equation}
with
\begin{equation}
k_{1}^{3} = 3\pi^{2}\rho_{c}
\label{b}
\end{equation}
\begin{equation}
k_{2}^{3}-k_{1}^{3} = 3\pi^{2}\rho_{m}
\label{b1}
\end{equation}
\begin{equation}
k_{3}^{3}-k_{2}^{3} = 3\pi^{2}\rho_{s}
\label{9}
\end{equation}
In Eq.~(\ref{a}) $\rho_{c}$ and $\rho_{s}$ are the core and shell electron 
density, and in Eq.~(\ref{b1}), $\rho_{m}$ is the density of the vacant orbitals
that lie between the core and the shell regions of occupied orbitals.

The exchange energy for the HEG that occupies the $k-space$ as described above
can be obtained exactly and is given as (MLDA stands for modified local-density
approximation)
\begin{equation}
E_{X}^{MLDA}=E_{X}^{core}+E_{X}^{shell}+E_{X}^{core-shell}
\label{10}
\end{equation}
where
\begin{equation}
E_{X}^{core} = V\left[-\frac{k_{1}^{4}}{4\pi^{3}}\right]
\label{11}
\end{equation}
is the exchange energy of the core electrons,
\begin{equation}
E_{X}^{shell} = -\frac{V}{8\pi^{3}}\left[2(k_{3}^{3}-k_{2}^{3})(k_{3}-k_{2})
+ (k_{3}^{2}-k_{2}^{2})^{2}\;ln\left(\frac{k_{3}+k_{2}}{k_{3}-k_{2}}\right)\right]
\label{12}
\end{equation}
is the exchange energy of the electrons in the shell, and
\begin{eqnarray}
E_{X}^{core-shell} & = & - \frac{V}{8 \pi^3} \big[2(k_3 - k_2) k_1^3+
2 (k_3^3 - k_2^3) k_1 + (k_2^2 - k_1^2)^2 \;ln
\left(\frac{k_2 + k_1}{k_2 - k_1} \right)  \nonumber  \\
 & & - (k_3^2 - k_1^2)^2 \;ln \left(\frac{k_3 + k_1}{k_3 - k_1} \right) \big]
\label{13}
\end{eqnarray}
represents the exchange energy of interaction between the core and the shell
electrons. Here $V$ is the volume of the HEG.  After adding the three terms, 
the exchange-energy can also be written in the form
\begin{eqnarray}
E_{X}^{MLDA} &=& \int\rho\left[\epsilon(k_{3}) - \epsilon(k_{2})
+\epsilon(k_{1})\right]d{\bf r} 
-\frac{1}{8\pi^{3}}\int\left[(k_{3}^{2}-k_{2}^{2})^{2}\;ln\left(\frac{k_{3}+k_{2}}{k_{3}-k_{2}}\right)\right]d{\bf r} \nonumber \\
& &-\frac{1}{8\pi^{3}}\int\left[(k_{2}^{2}-k_{1}^{2})^{2}\;ln\left(\frac{k_{2}+k_{1}}{k_{2}-k_{1}}\right)\right]d{\bf r} \nonumber \\ 
& &+\frac{1}{8\pi^{3}}\int\left[(k_{3}^{2}-k_{1}^{2})^{2}\;ln\left(\frac{k_{3}+k_{1}}{k_{3}-k_{1}}\right)\right]d{\bf r} \nonumber \\
\label{14}
\end{eqnarray}
where $\epsilon(k_f)$ represents the exchange-energy per particle when the HEG is in 
its ground-state with the Fermi momentum equal to $k_f$. The equation above has a nice 
interpretation:  The integral on the right-hand side represents the exchange energy of 
the system of electrons with density $\rho$ when per electron energy is written as 
$[\epsilon(k_{3}) - \epsilon(k_{2}) +\epsilon(k_{1})]$, i.e. the per electron energy is 
given according to the occupation in the $k-space$ (compare with Eq.~(\ref{10})). The $log.$ 
terms, on the other hand, have no such simple interpretation. They have the kinetic 
energy density in them but we have not been able to write the terms in as easy a form 
as the first term. That the functional above has all the right limits if we take 
$k_{1}=k_{2}$ or $k_{2}=k_{3}$ is easily verified. Finally, the modified local-spin 
density (LSD) functional $E_{X}^{MLSD}[\rho_{\alpha},\rho_{\beta}]$ in terms of the 
spin densities $\rho_{\alpha}$ and $\rho_{\beta}$ is easily obtained from the 
functional above as
\begin{equation}
E_{X}^{MLSD}[\rho_{\alpha},\rho_{\beta}] = \frac{1}{2}E_{X}^{MLDA}
[2\rho^{\alpha}] + \frac{1}{2}E_{X}^{MLDA}[2\rho^{\beta}] 
\label{lsda}
\end{equation}

Having derived the exchange functional for the HEG, we now apply it to the 
excited-states of various atoms to check if the functional above gives exchange
energy differences accurately. The excited-states chosen are such that they can be 
represented by a single Slater determinant so that the LDA is expected to be a good 
approximation \cite{ziegler,vonB} for them.  The different radii in the $k-space$
($k_{1}$, $k_{2}$ and $k_{3}$) needed to evaluate the exchange energy are found by 
Eqs.~(\ref{b}), (\ref{b1}) and (\ref{9}). For each state (ground and excited), the same set
of orbitals \cite{fnote1}  is employed to get the Hartree-Fock and the LSD exchange 
energies. We calculate the LSD and MLSD exchange energies using spherical spin 
densities since the effect of non-sphericity on the total exchange energy should be 
small \cite{janak}.This is because of the fact that in the Levy-Nagy formalism 
\cite{levy1}, the excited-state energy is obtained through variational minimization. 
Therefore inclusion of non-sphericity in the density would not cause as large a change 
in the total excited-state energy as the use of an appropriate exchange energy 
functional.  Indeed the results for the lowest lying multiplets also indicate this 
\cite{janak}.

In Table 1, we show the difference between the excited-state exchange energy and the
ground-state exchange energy for some atoms and ions. In the first column we give the 
difference as obtained by the Hartree-Fock expression for the exchange energy.  In the 
second column, the numbers are given for both the excited-state and the ground-state 
exchange energies obtained by employing the ground-state LSD functional.  The third 
column gives the exchange energy difference when the excited-state exchange energy
is calculated using the functional of Eq.~(\ref{lsda}). It is clearly seen that the 
ground-state LSD approximation underestimates this energy difference.  This is
not surprising since the ground-state functional would give a larger exchange energy
for the excited-state than what a proper excited-state functional should give. However,
when the functional of Eq.~(\ref{lsda}) is employed to calculate the exchange energy for 
the excited-states we find, to our surprise, that for the majority of the atoms the 
functional overestimates the differences by a large amount, whereas we expected to 
find the error to be about $10\%$ which is the general LDA exchange energy error. We 
note that this large difference cannot come because we have spherical densities. If 
non-spherical densities are used, the difference may increase even further.  For 
example, for the fluorine atom, the ground-state exchange energy will become more 
negative for non-spherical densities. On the other hand, the excited-state exchange 
energy will remain unchanged since the density is already spherical.  This will result 
in an even larger difference in the exchange energies of the two states.
 
We now look for possible sources of error in the exchange-energy differences when
the functional of Eq.~(\ref{lsda}) is employed to get the exchange energy for the
excited-states. For this we examine Eq.~(\ref{7}) in which the last four terms on the
right hand side represent the exchange energy difference. Thus
\begin{eqnarray}
\Delta E_{X} & = &{\sum_{j(\sigma_j=\sigma_{rem})}} \left<\phi_{rem}{({\bf r}_{1})}
     \phi_{j}{({\bf r}_{2}) } \left|\frac{1}{ |{\bf r}_{1}-{\bf r}_{2}|} \right|\phi_{j}
     {({\bf r}_{1})}\phi_{rem}{({\bf r}_{2})} \right> \nonumber \\
          &   & -\frac{1}{2}\int\int\frac{|\phi_{rem}({\bf r}_{1})|^{2}|\phi_{rem}
({\bf r}_{2})|^{2}}{|{\bf r}_{1}-{\bf r}_{2}|}d{\bf r}_{1}d{\bf r}_{2} \nonumber \\
 & &-\frac{1}{2}\int\int\frac{|\phi_{add}({\bf r}_{1})|^{2}|\phi_{add}
({\bf r}_{2})|^{2}}{|{\bf r}_{1}-{\bf r}_{2}|}d{\bf r}_{1}d{\bf r}_{2} \nonumber \\
                &   & - {\sum_{j(j\neq add)(\sigma_j=\sigma_{add})}} \left<\phi_{add}
{({\bf r}_{1})}
     \phi_{j}{({\bf r}_{2}) } \left|\frac{1}{ |{\bf r}_{1}-{\bf r}_{2}|} \right|\phi_{j}
     {({\bf r}_{1})}\phi_{add}{({\bf r}_{2})} \right>\;
\label{15}
\end{eqnarray}
It is the LDA values to this term that are given in Table 1. The source of error in 
this term we suspect is the LDA treatment of the self-exchange energies of the orbitals
$\phi_{rem}$ and $\phi_{add}$ involved in the electron transfer. To make the functional 
more accurate we make the self-interaction correction (SIC) for both these orbitals.  
This is done by subtracting \cite{perdewz}
\begin{equation}
E_{X}^{SIC}[\phi] = \frac{1}{2}\int\int\frac{|\phi({\bf r}_{1})|^{2}
   |\phi({\bf r}_{2})|^{2})}{|{\bf r}_{1}-{\bf r}_{2}|}d{\bf r}_{1}d{\bf r}_{2}
   + E_{X}^{LSD}[\rho(\phi)]\;,
\label{16}
\end{equation}
where $\rho(\phi)$ is the orbital density for the orbital $\phi$, from the 
$E^{MLSD}_{X}$ functional.  Thus the final expression for the exchange-energy that 
we have is
\begin{equation}
E_X^{MLSDSIC} = E_X^{MLSD}-E_{X}^{SIC}[\phi_{rem}]-E_{X}^{SIC}[\phi_{add}]\;
\label{18}
\end{equation}
This gives the exchange energy difference between the excited-state and the
ground-state to be
\begin{equation}
\Delta E_{X} = E^{MLSD}_{X}[\rho_{excited}]-E^{LSD}_{X}[\rho_{ground}]
-E^{SIC}_{X}[\phi_{rem}]-E^{SIC}_{X}[\phi_{add}]\;
\label{18a}
\end{equation}
We have also computed the exchange energy differences given by the functional in 
Eq.~(\ref{18}) and shown them in Table 1. As is evident from the numbers displayed 
there, the functional of Eq.~(\ref{18}) gives highly accurate exchange-energy 
differences for all the systems considered.  When the exchange-energy difference 
between the ground- and the excited-state is small, the HF, LSD and the functionals 
derived above, all give roughly the same results.  However, when this difference 
is large, the LDA underestimates the magnitude of the difference by a large amount 
whereas the functional of Eq.~(\ref{lsda}) overestimates it. Only when the latter is 
corrected for the self-interaction then the difference is almost the same as the 
Hartree-Fock difference. Notice that SIC is made only for the orbitals involved in the 
transition. Thus despite this explicit orbital dependence, the functional is still
quite simple and easy to deal with. 

Making the self-interaction correction may deceptively lead the reader to momentarily 
think that our approach may be nothing more than treating the exchange energy within
the SICLDA approach for both the ground and the excited-states. However, this is
not so.  If the SICLDA exchange energy functional is used, the difference between
the exchange energies for the two states would be (for keeping the expression simple, 
we are using the same set of orbitals for the two states):
\begin{equation}
\Delta E_{X} = E^{LSD}_{X}[\rho_{excited}]-E^{LSD}_{X}[\rho_{ground}]
+E^{SIC}_{X}[\phi_{rem}]-E^{SIC}_{X}[\phi_{add}]\;
\label{16a}
\end{equation}
Expressions in Eqs.~(\ref{18a}) and (\ref{16a}) differ in two significant ways: First,
the exchange energy functionals used for the ground and excited states are different, 
and secondly $E^{SIC}_{X}[\phi_{rem}]$ is subtracted in Eq.~(\ref{18a}) whereas it is 
added in Eq.~(\ref{16a}).  A careful look at Eq.~(\ref{16a}) also indicates that the 
excited-state energies in SICLDA scheme should not come out to be any different from 
those obtained from the LDA calculations, because $E^{SIC}$ for the two orbitals 
involved in the transition would tend to cancel. This is what has been observed in the 
past \cite{GunnarJ,perdewz,HarrisJ}.

We note that we do have a choice of writing the first and the second terms in 
Eq.~(\ref{15}) as 
\begin{eqnarray}
 {\sum_{j(j\neq rem)(\sigma_j=\sigma_{rem})}} \left<\phi_{rem}{({\bf r}_{1})}
     \phi_{j}{({\bf r}_{2}) } \left|\frac{1}{ |{\bf r}_{1}-{\bf r}_{2}|} \right|\phi_{j}
     {({\bf r}_{1})}\phi_{rem}{({\bf r}_{2})} \right> \nonumber \\  
     + \frac{1}{2}\int\frac{|\phi_{rem}({\bf r}_{1})|^{2}|\phi_{rem}
({\bf r}_{2})|^{2}}{|{\bf r}_{1}-{\bf r}_{2}|}d{\bf r}_{1}d{\bf r}_{2} 
\label{17}
\end{eqnarray}
and then make the self-interaction correction for the orbital $\phi_{rem}$.  However,
that would bring $E^{SIC}_{X}[\phi_{rem}]$ with a positive sign in $E_{X}^{MLSDSIC}$,
and the resulting functional will not be as accurate. Although we do not fully 
understand why this happens, we now give a qualitative argument as to why the 
functional of Eq.~(\ref{18}) gives accurate exchange energy differences. We feel that
the LDA should be reasonably accurate when the integral over $k$ is continuous. As 
written in Eq.~(\ref{15}), the sum in the first term is continuous except for the
exchange term involving $\phi_{rem}$ and $\phi_{add}$.  Thus the LDA to the first term 
should be reasonably accurate. This brings in the self-interaction energy of the 
electron removed with a negative sign in front.  By including the self-interaction 
correction for the removed electron only, we find that the error in the exchange energy
difference reduces to about $10\%$ of the corresponding HF value. To make the 
difference even more accurate, we now consider the term for the orbital $\phi_{add}$ 
where the electron is added.  There, when the LDA is made, the electron comes in with 
its self-interaction so for the added orbital too $E_{X}^{SIC}$ should be subtracted 
to make the results for the energy difference comparable to the Hartree-Fock results.

Having obtained the functional to calculate accurate exchange energy difference, we now
apply it to a large number of excited-states of the class considered here and find
that we get the transition energies very close to those given by the Hartree-Fock 
theory.

\section{Results}
We employ the exchange functional $E_{X}^{MLSDSIC}$ proposed above to obtain the 
transition energies for a variety of excitations in different atoms. We find that for 
all the systems the transition energies obtained by us are very close to the
corresponding Hartree-Fock energies \cite{fnote2}.  Our calculations proceed as 
follows: (a) We get the ground-state energy by solving the Kohn-Sham equation with the 
effective exchange potential calculated using the Dirac formula \cite{dirac}.
(b)  We then solve the Kohn-Sham equation with the same (corresponding to the 
ground-state) functional for the excited-state configuration.  This gives us the 
excited-state energy $E_{LSD}$, and the LSD exchange energy $E_{X}^{LSD}$ for the
excited-state. The difference between $E_{LSD}$ and the ground-state 
energy $E_{0}$ gives us the transition energy $\Delta E_{LSD}$. (c) We then employ the 
Kohn-Sham orbitals from the excited-state LSD calculation to get the modified LSD 
exchange energy including SIC by employing the functional $E_{X}^{MLSDSIC}$ of 
Eq.~(\ref{18}). (d) The modified transition energy $\Delta E_{MLSDSIC}$ is then given
as
\begin{equation}
\Delta E_{MLSDSIC} = \Delta E_{LSD} + E_{X}^{MLSDSIC} - E_{X}^{LSD} 
\label{17a}
\end{equation}
Although we have not performed self-consistent calculations with the new functional, 
self-consistency is not expected to affect the results significantly. This is because, 
as we shall see in the results, the major difference in the transition energies given 
by different functionals arises from the difference in the value of the exchange energy 
itself. We also compare our results with the transition energies obtained by the 
exchange-only time-dependent density-funcitional theory (TDDFT) applied within the 
single-pole approximation.  We find that our results are comparable to the TDDFT 
results. In the following, we have considered three different cases of electron 
transfer: electron making a transition from an `s' to a `p' orbital; from an `s' to a 
`d' orbital and from a `p' to a `d' orbital.

\subsection{Electron transfer from an `s' to a `p' orbital}
In this section we consider the cases when one or two electrons are transferred
from an inner $s$ orbital to an outer $p$ orbital.  Shown in Table 2 are the 
transition energies $\Delta E_{HF}$, $\Delta E_{LSD}$ and $\Delta E_{MLSDSIC}$ for 
some light atoms and ions when one of their inner electrons is excited to the lowest 
available orbital. The excitation energy in these systems is such that for some of 
them $\Delta E_{LSD}$ is close to $\Delta E_{HF}$ but for others it is not.  However, 
$\Delta E_{MLSDSIC}$ is uniformly accurate for all the systems. We note that the error 
in $\Delta E_{LSD}$ is almost fully from the error in the corresponding exchange energy 
difference.  This is evident from a comparison of the numbers in Table 1 (for the 
exchange energy differences) and in Table 2. Thus major difference in  $\Delta E$ 
comes from the error in calculating the exchange energy. As noted earlier, 
self-consistency effects are much smaller compared to the differences arising from the 
use of the ground-state exchange energy functional for the excited-state also.  
Our results also match well with, and in some cases are better than, the TDDFT results 
shown in the last column of the table.

In Table 3, we look at the excitation energies of the alkali atoms and $Mg^{+}$
by exciting an electron from the uppermost orbital to an outer orbital.  These
are weakly bound systems and as such their excitation energies are relatively
smaller.  Thus they provide a good testing ground for the proposed functional.
An interesting point about these systems is that the LSD itself gives excitation
energies close to the HF excitation energies. It is therefore quite gratifying to
see that the transition energies obtained by the new functional also 
are of very good quality, although the present method tends to slightly overestimate the
transition energies. The TDDFT method also gives similar numbers although it
overestimates the transition energies by a slightly larger amount.

Next we consider some bigger atoms where we can excite the electron from
more than one inner orbital.  Shown in Tables 4 and 5 are the excitation energies
for the atoms in the third row of the periodic table. In Table 4, we consider
an electron being excited from the $3s$ orbital to the $3p$ orbital. In all
these case $\Delta E_{LSD}$ is smaller than the true energy difference
whereas the present functional gives highly accurate estimates of the transition
energy. Notice again that the error in the value of $\Delta E_{LSD}$ arises 
mainly from the error in the exchange energy. The TDDFT results in these cases
too are of quality comparable to the present method.

In Table 5, we show the transition energies for the same set of atoms and ions
as in Table 4, but for the electron now being excited from the $2s$ orbital
to the $3p$ orbital.  Consequently the energy of excitation is much larger
in this case. The LSD in all these cases underestimates the excitation energy,
whereas the present functional gives accurate results although slightly
overestimating them.  However, the error with respect to the LSD is reduced by
a factor of $5$ or more. Thus the proposed functional is accurate for transitions 
from a shallow level as well as from a deep level. We find that the TDDFT results
in the present case are not as accurate as in the cases studied above. 

Shown in Table 6 are the excitation energies for a group of atoms for which
the LSD gives transition energies very close to the HF excitation energies.
In all the cases we find that the functional proposed here is able to give
accurate excitation energies.  Thus we find that when the LSD results are
accurate, so are the results given by the new functional.  What is significant,
however, is that when the LSD results are poor, the new functional properly
corrects the error in the LSD.

Finally, we consider the cases where two electrons are excited to the higher orbitals. 
In this case the functional $E_{X}^{MLSDSIC}$ is evaluated by subtracting the
SIC energy from the $E_{X}^{MLSD}$ for both the electrons.
As already pointed out, double excitations are difficult to deal with in the TDDFT 
approach to finding excitation energies, because the theory is based on the first-order
perturbation theory of non-interacting particles. Results for different excitations for
a variety of atomic systems are shown in Table 7.  As is evident from the table, for 
all the systems, our method gives excellent results whereas the LSD underestimates the 
energies. In the case of double excitations, no comparison with the TDDFT results
can be made because a satisfactory TDDFT of double excitations does not exist.

In all the cases above, we have compared our results with those of Hartree-Fock
theory and those obtained from exchange-only TDDFT.  We do so because in our
work we have not taken into account the effect of correlations.  We note that
although in atoms Hartree-Fock theory gives total energies which are very close 
to the experimental energies, correlation effects become relatively more important 
in calculating transition energies which is the difference between total energies.
Thus a comparison with experimental transition energies would be meaningful only
after correlation effects are properly taken into account.

\subsection{Electron transfer from an `s' or a `p' orbital to a `d' orbital}
In this section we consider the case when a $3s$ or a $3p$ electron is transferred 
to an incompletely filled $3d$ orbital in transition metals $Sc, Ti, V, Mn, Fe, Co$
and $Ni$. The results for our calculations on these systems are shown in Tables 8
and 9.

In Table 8, numbers are shown for transitions from the $3s$ orbital of these atoms
to their $3d$ orbital.  As is clear from the table, whereas the LSD results 
underestimate the transfer energies, our results compare well with those of Hartree-Fock
theory.  On the other hand, we find that the TDDFT results also underestimate the
transition energies in comparison to the Hartree-Fock theory results.

In Table 9, results for the transition $3p$ $\rightarrow$ $3d$ are shown. Since there are 
three down spin electrons in the $3p$ orbital, two electrons are left in it after an
electron is excited to the $3d$ orbital. The question arises whether to treat these
electrons as core electrons, as shell electrons or divide them in the core and the
shell; the three possibilities give different answers.  When the electrons are treated 
as the core electrons, the transition energy comes out to be the smallest and it
is the largest when they are treated as the shell electrons.  We have taken the 
smallest transition energy obtained by us and compared it to the energy of transition 
to the largest possible L value (so the smallest transition energy) excited-state for a 
given spin. In these cases too, we find that for $Sc, Ti$ and $V$, our method gives
results which are better than the LSD results, but for systems with more than half
filled $d-shell$, our method overestimates the transition energy whereas the LSD 
underestimates it. The three cases of $Fe, Co$ and $Ni$ are where our results do not
match with those of Hartree-Fock theory.

A tough problem in calculating transfer energies is that of electron transferring
from the $4s$ orbital to the $3d$ orbital in the transition metals considered above.
The problem has been well investigated \cite{GunnarJ, HarrisJ} in the past and as in 
all the cases considered so far, LSD underestimates these energies by large amount.  
We have applied our functional to obtain these transfer energies to see if we could get 
the correct answer. However, for these $s \rightarrow d$ transfer energies, our method 
gives hardly any improvement over the LSD results; in fact for most of the systems, we 
get a transition energy which is lower than the LSD energy.  Furher investigations of 
this problem are being made.

\section{Discussion and concluding remarks}
In the above we have presented a new LDA-like functional for obtaining the excitation
energies.  It has been employed to investigate over $50$ excited states. The results
show that our procedure gives accurate excitation energies for all of them, whereas for 
most of the systems the LSD underestimates the energy difference. Thus we have 
demonstrated that it is possible to construct excited-state exchange energy functionals 
that are capable of giving transition energies close to the exact theory.  We have 
worked within the exchange-only approximation and have chosen a particular class of 
excited-states. What we have learnt through the study reported here is that there exist
exchange energy functionals of densities that are more accurate than the ground-state
functional.  However the structure of these functionals is not as simple as
the ground-state functional. 

In Levy-Nagy theory \cite{levy1}, while defining the bi-density functional for the 
excited-states through constrained search formulation \cite{levy2} (Eq.~(\ref{2})), the 
wavefunctions involved in the minimization procedure are those which are orthogonal to 
the lower energy wavefunctions. The latter are supposed to be determined by the 
ground-state density $\rho_0(\bf r)$.  In constructing our functional, by looking at 
the orbital occupation in the excited-states, we occupy the $k-space$ in a similar 
manner, representing the unoccupied orbitals by a gap in it. Thus in our case the 
orthogonality condition described above is taken care of, to a large extent, by the 
gap in the $k-space$.  This also reflects an implicit dependence on the ground-state 
density.

We are now working on functionals for states other than those considered
in this paper.  As pointed out in the introduction, excited-state functionals are
not universal and therefore have to be dealt with separately for different kinds
of excited-states. A fundamental question that still remains unanswered is why is
it that representing an excited-state by the corresponding excited-state of HEG
does not by itself give at least as accurate exchange energies for the excited-states 
that the LDA does for the ground-state. We do not yet have a satisfactory answer
for this.  What is clear however is that a combination of this functional and 
appropriate SIC gives highly accurate results. In this work, we have also not looked 
at the correlation energy functionals.  Can correlation energy functionals be 
developed along similar lines?  We trust that it should be possible and are working on 
this problem.

{\bf Acknowledgement:} We thank Professor K.D. Sen for providing the Hartree-Fock
data on excited-states of atoms. Comments of Rajan Pandey on the manuscript are
appreciated.

\newpage
\begin{table}
\caption{Difference in the exchange energies of the ground- and excited-states of 
some atoms and ions.  The First column gives the atom/ion and the transition, the 
second column the difference $\Delta E_X^{HF}$ as obtained in Hartree-Fock theory, the 
third column the difference $\Delta E_X^{LSD}$ given by the ground-state energy 
functional.  The fourth and the fifth column describes the difference as obtained with the 
functional proposed in this paper.  The fourth column gives the exchange-energy difference 
$\Delta E_X^{MLSD}$ obtained by employing the functional of Eq.~(\ref{lsda}) whereas the fifth
column gives that given by the functional of Eq.~(\ref{18}), $\Delta E_X^{MLSDSIC}$. Numbers
given are in atomic units.}
\vspace{0.2in}
\begin{tabular}{lcccc}
\hline
atoms/ions&$\Delta$$E_X^{HF}$&$\Delta$$E_X^{LSD}$&$\Delta$$E_X^{MLSD}$&
$\Delta$$E_X^{MLSDSIC}$\\
\hline
$Li(2s^{1}\;^{2}S\rightarrow2p^{1}\;^{2}P)$ &0.0278&0.0264&0.0587&0.0282 \\
$B(2s^{2}2p^{1}\;^{2}P\rightarrow2s^{1}2p^{2}\;^{2}D)$ &0.0353&0.0319&0.0998&0.0412 \\
$C(2s^{2}2p^{2}\;^{3}P\rightarrow2s^{1}2p^{3}\;^{3}D)$ &0.0372&0.0332&0.1188&0.0454 \\
$N(2s^{2}2p^{3}\;^{4}S\rightarrow2s^{1}2p^{4}\;^{4}P)$ &0.0399&0.0353&0.1381&0.0503 \\
$O(2s^{2}2p^{4}\;^{3}P\rightarrow2s^{1}2p^{5}\;^{3}P)$ &0.1582&0.0585&0.2634&0.1624 \\
$F(2s^{2}2p^{5}\;^{2}P\rightarrow2s^{1}2p^{6}\;^{2}S)$ &0.3021&0.0891&0.3908&0.2765 \\
$Ne^{+}(2s^{2}2p^{5}\;^{2}P\rightarrow2s^{1}2p^{6}\;^{2}S)$ &0.3339&0.0722&0.4397&0.3037 \\
$S(3s^{2}3p^{4}\;^{3}P\rightarrow3s^{1}3p^{5}\;^{3}P)$ &0.1106&0.0475&0.1798&0.1252 \\
$Cl^{+}(3s^{2}3p^{4}\;^{3}P\rightarrow3s^{1}3p^{5}\;^{3}P)$ &0.1257&0.0483&0.2050&0.1441\\
$Cl(3s^{2}3p^{5}\;^{2}P\rightarrow3s^{1}3p^{6}\;^{2}S)$ &0.2010&0.0603&0.2567&0.1969\\
\hline
\end{tabular}
\end{table}

\begin{table}
\caption{Transition energies, in atomic units, of an electron being excited from the 
$2s$ orbital of some atoms to their $2p$ orbital.  The first column gives this energy
as obtained in Hartree-Fock theory.  The numbers in the second column are obtained by 
employing the ground-state LDA for both the ground- and the excited-state.  The last 
column gives the energies given by employing the ground-state LDA for the ground-state 
and the functional of Eq.~(\ref{18}) for the excited-state.}
\vspace{0.2in}
\begin{tabular}{lcccc}
\hline
atoms/ions & $\Delta$$E_{HF}$ & $\Delta$$E_{LSD}$ &   $\Delta$$E_{MLSDSIC}$ &
$\Delta$$E_{TDDFT}$  \\
\hline
$N(2s^{2}2p^{3}\;^{4}S\rightarrow2s^{1}2p^{4}\;^{4}P)$ &0.4127&0.3905&0.4014&0.4153 \\
$O^{+}(2s^{2}2p^{3}\;^{4}S\rightarrow2s^{1}2p^{4}\;^{4}P)$ &0.5530&0.5397&0.5571&0.5694 \\
$O(2s^{2}2p^{4}\;^{3}P\rightarrow2s^{1}2p^{5}\;^{3}P)$ &0.6255&0.5243&0.6214&0.5912 \\
$F^{+}(2s^{2}2p^{4}\;^{3}P\rightarrow2s^{1}2p^{5}\;^{3}P)$ &0.7988&0.6789&0.8005&0.7651 \\
$F(2s^{2}2p^{5}\;^{2}P\rightarrow2s^{1}2p^{6}\;^{2}S)$ &0.8781&0.6671&0.8573&0.7659 \\
$Ne^{+}(2s^{2}2p^{5}\;^{2}P\rightarrow2s^{1}2p^{6}\;^{2}S)$ &1.0830&0.8334&1.0607&0.9546 \\
\hline
\end{tabular}
\end{table}

\begin{table}
\caption{The caption is the same as that for Table 2 except that we are now
considering transitions from the outermost orbital to an upper orbital for
weakly bound systems.}
\vspace{0.2in}
\begin{tabular}{lcccc}
\hline
atoms/ions & $\Delta$$E_{HF}$ & $\Delta$$E_{LSD}$ &   $\Delta$$E_{MLSDSIC}$ &
$\Delta$$E_{TDDFT}$  \\
\hline
$Li(2s^{1}\;^{2}S\rightarrow2p^{1}\;^{2}P)$ &0.0677&0.0646&0.0672&0.0724 \\
$Na(3s^{1}\;^{2}S\rightarrow3p^{1}\;^{2}P)$ &0.0725&0.0751&0.0753&0.0791 \\
$Mg^{+}(3s^{1}\;^{2}S\rightarrow3p^{1}\;^{2}P)$ &0.1578&0.1585&0.1696&0.1734 \\
$K(4s^{1}\;^{2}S\rightarrow4p^{1}\;^{2}P)$ &0.0516&0.0556&0.0580&0.0608 \\
\hline
\end{tabular}
\end{table}

\begin{table}
\caption{Electron transition energy from the $3s$ to the $3p$ orbital in some atoms.}
\vspace{0.2in}
\begin{tabular}{lcccc}
\hline
atoms/ions & $\Delta$$E_{HF}$ & $\Delta$$E_{LSD}$ &   $\Delta$$E_{MLSDSIC}$ &
$\Delta$$E_{TDDFT}$  \\
\hline
$P(3s^{2}3p^{3}\;^{4}S\rightarrow3s^{1}3p^{4}\;^{4}P)$ &0.3023&0.2934&0.3055&0.3183 \\
$S(3s^{2}3p^{4}\;^{3}P\rightarrow3s^{1}3p^{5}\;^{3}P)$ &0.4264&0.3615&0.4334&0.4122 \\
$Cl^{+}(3s^{2}3p^{4}\;^{3}P\rightarrow3s^{1}3p^{5}\;^{3}P)$ &0.5264&0.4482&0.5403&0.5113 \\
$Cl(3s^{2}3p^{5}\;^{2}P\rightarrow3s^{1}3p^{6}\;^{2}S)$ &0.5653&0.4301&0.5630&0.4996 \\
$Ar^{+}(3s^{2}3p^{5}\;^{2}P\rightarrow3s^{1}3p^{6}\;^{2}S)$ &0.6769&0.5174&0.6766&0.6007 \\
\hline
\end{tabular}
\end{table}

\begin{table}
\caption{Electron transition energy from the $2s$ to the $3p$ orbital in the
same atoms as in Table 4.}
\vspace{0.2in}
\begin{tabular}{lcccc}
\hline
atoms/ions & $\Delta$$E_{HF}$ & $\Delta$$E_{LSD}$ &   $\Delta$$E_{MLSDSIC}$ &
$\Delta$$E_{TDDFT}$  \\
\hline
$P(2s^{2}3p^{3}\;^{4}S\rightarrow2s^{1}3p^{4}\;^{4}P)$ &6.8820&6.4188&6.9564&6.1573 \\
$S(2s^{2}3p^{4}\;^{3}P\rightarrow2s^{1}3p^{5}\;^{3}P)$ &8.2456&7.7337&8.3271&7.4533 \\
$Cl^{+}(2s^{2}3p^{4}\;^{3}P\rightarrow2s^{1}3p^{5}\;^{3}P)$ &9.8117&9.2551&9.8997&8.9618 \\
$Cl(2s^{2}3p^{5}\;^{2}P\rightarrow2s^{1}3p^{6}\;^{2}S)$ &9.7143&9.1653&9.8171&8.8686 \\
$Ar^{+}(2s^{2}3p^{5}\;^{2}P\rightarrow2s^{1}3p^{6}\;^{2}S)$ &11.3926&10.8009&11.5061&10.4901 \\
\hline
\end{tabular}
\end{table}

\begin{table}
\caption{Electron transition energy when the upper state is not the lowest
energy multiplet.} 
\vspace{0.2in}
\begin{tabular}{lcccc}
\hline
atoms/ions & $\Delta$$E_{HF}$ & $\Delta$$E_{LSD}$ &   $\Delta$$E_{MLSDSIC}$ &
$\Delta$$E_{TDDFT}$  \\
\hline
$B(2s^{2}2p^{1}\;^{2}P\rightarrow2s^{1}2p^{2}\;^{2}D)$ &0.2172&0.1993&0.2061&0.2168 \\
$C^{+}(2s^{2}2p^{1}\;^{2}P\rightarrow2s^{1}2p^{2}\;^{2}D)$ &0.3290&0.3078&0.3216&0.3325 \\
$C(2s^{2}2p^{2}\;^{3}P\rightarrow2s^{1}2p^{3}\;^{3}D)$ &0.2942&0.2878&0.2967&0.3090 \\
$N^{+}(2s^{2}2p^{2}\;^{3}P\rightarrow2s^{1}2p^{3}\;^{3}D)$ &0.4140&0.4149&0.4305&0.4433 \\
$Si^{+}(3s^{2}3p^{1}\;^{2}P\rightarrow3s^{1}3p^{2}\;^{2}D)$ &0.2743&0.2632&0.2799&0.2864 \\
$Si(3s^{2}3p^{2}\;^{3}P\rightarrow3s^{1}3p^{3}\;^{3}D)$ &0.2343&0.2356&0.2442&0.2567 \\
\hline
\end{tabular}
\end{table}

\begin{table}
\caption{Excitation energies of some atoms when two electrons are excited.}
\vspace{0.2in}
\begin{tabular}{lccc}
\hline
atoms/ions & $\Delta$$E_{HF}$ & $\Delta$$E_{LSD}$ &   $\Delta$$E_{MLSDSIC}$  \\
\hline
$Be(2s^{2}\;^{1}S\rightarrow2p^{2}\;^{1}D)$ &0.2718&0.2538&0.2655 \\
$B(2s^{2}2p^{1}\;^{2}P\rightarrow2p^{3}\;^{2}D)$ &0.4698&0.4117&0.4798 \\
$C^{+}(2s^{2}2p^{1}\;^{2}P\rightarrow2p^{3}\;^{2}D)$ &0.6966&0.6211&0.7180 \\
$C(2s^{2}2p^{2}\;^{3}P\rightarrow2p^{4}\;^{3}P)$ &0.7427&0.5950&0.7312 \\
$N^{+}(2s^{2}2p^{2}\;^{3}P\rightarrow2p^{4}\;^{3}P)$ &1.0234&0.8369&1.0143 \\
$N(2s^{2}2p^{3}\;^{4}S\rightarrow2p^{5}\;^{2}P)$ &1.1789&0.9440&1.1785 \\
$O^{+}(2s^{2}2p^{3}\;^{4}S\rightarrow2p^{5}\;^{2}P)$ &1.5444&1.2552&1.5480 \\
$O(2s^{2}2p^{4}\;^{3}P\rightarrow2p^{6}\;^{1}S)$ &1.5032&1.1333&1.4736 \\
$F^{+}(2s^{2}2p^{4}\;^{3}P\rightarrow2p^{6}\;^{1}S)$ &1.8983&1.4381&1.8494 \\
$Mg(3s^{2}\;^{1}S\rightarrow3p^{2}\;^{1}D)$ &0.2578&0.2555&0.2651 \\
$S(3s^{2}3p^{4}\;^{3}P\rightarrow3p^{6}\;^{1}S)$ &1.0273&0.7807&1.0266 \\
$P(3s^{2}3p^{3}\;^{4}S\rightarrow3p^{5}\;^{2}P)$ &0.8539&0.6927&0.8680 \\
$Si^{+}(3s^{2}3p^{1}\;^{2}P\rightarrow3p^{3}\;^{2}D)$ &0.5856&0.5377&0.6230 \\
$Si(3s^{2}3p^{2}\;^{3}P\rightarrow3p^{4}\;^{3}P)$ &0.5860&0.4928&0.5986 \\
$Cl^{+}(3s^{2}3p^{2}\;^{3}P\rightarrow3p^{4}\;^{3}P)$ &1.2535&0.9551&1.2516 \\
\hline
\end{tabular}
\end{table}

\begin{table}
\caption{Electron transition energy when an `s' electron is transferred to
a `d' orbital.}
\vspace{0.2in}
\begin{tabular}{lcccc}
\hline
atoms/ions & $\Delta$$E_{HF}$ & $\Delta$$E_{LSD}$ &   $\Delta$$E_{MLSDSIC}$ &
$\Delta$$E_{TDDFT}$  \\
\hline
$Sc(3s^{2}3d^{1}\;^{2}D\rightarrow3s^{1}3d^{2}\;^{2}G)$ &2.1562&1.8584&2.1223&1.8649 \\
$Ti(3s^{2}3d^{2}\;^{3}F\rightarrow3s^{1}3d^{3}\;^{5}F)$ &2.2453&1.9740 &2.2061&-----  \\
$Ti(3s^{2}3d^{2}\;^{3}F\rightarrow3s^{1}3d^{3}\;^{3}H)$ &2.3861&2.0827 &2.3649&2.0951  \\
$V(3s^{2}3d^{3}\;^{4}F\rightarrow3s^{1}3d^{4}\;^{4}H)$ &2.6098&2.3107 &2.6106&2.3266 \\
$Mn(3s^{2}3d^{5}\;^{6}S\rightarrow3s^{1}3d^{6}\;^{6}D)$ &3.1331&2.7860 &3.1199&2.8062  \\
$Fe(3s^{2}3d^{6}\;^{5}D\rightarrow3s^{1}3d^{7}\;^{5}F)$ &3.4187&3.0483 &3.4527&3.0755  \\
$Co(3s^{2}3d^{7}\;^{4}F\rightarrow3s^{1}3d^{8}\;^{4}F)$ &3.7623&3.3178 &3.7955&3.3516  \\
$Ni(3s^{2}3d^{8}\;^{3}F\rightarrow3s^{1}3d^{9}\;^{3}D)$ &4.1204&3.5949 &4.1476&3.6351 \\
\hline
\end{tabular}
\end{table}

\begin{table}
\caption{Electron transition energy when a `p' electron is transferred to
a `d' orbital.}
\vspace{0.2in}
\begin{tabular}{lcccc}
\hline
atoms/ions & $\Delta$$E_{HF}$ & $\Delta$$E_{LSD}$ &   $\Delta$$E_{MLSDSIC}$ &
$\Delta$$E_{TDDFT}$  \\
\hline
$Sc(3p^{6}3d^{1}\;^{2}D\rightarrow3p^{5}3d^{2}\;^{2}H)$ &1.1295&1.1018 &1.1245&1.2128  \\
$Ti(3p^{6}3d^{2}\;^{3}F\rightarrow3p^{5}3d^{3}\;^{3}I)$ &1.2698&1.2478 &1.2728&1.3586  \\
$V(3p^{6}3d^{3}\;^{4}F\rightarrow3p^{5}3d^{4}\;^{4}I)$ &1.4153&1.3959 &1.4227&1.5042  \\
$Mn(3p^{6}3d^{5}\;^{6}S\rightarrow3p^{5}3d^{6}\;^{6}F)$ &1.7270&1.6431 &1.6726&1.8073  \\
$Fe(3p^{6}3d^{6}\;^{5}D\rightarrow3p^{5}3d^{7}\;^{5}G)$ &1.8785&1.8784 &2.0061&1.9898  \\
$Co(3p^{6}3d^{7}\;^{4}F\rightarrow3p^{5}3d^{8}\;^{4}G)$ &2.1178&2.0568 &2.2778&2.1755  \\
$Ni(3p^{6}3d^{8}\;^{3}F\rightarrow3p^{5}3d^{9}\;^{3}F)$ &2.4232&2.2402 &2.5518&2.3656  \\
\hline
\end{tabular}
\end{table}


\begin{thebibliography}{unsrt}
\bibitem{fnote2} The energies are calculated by solving the Kohn-Sham equation
with the Harbola-Sahni potential. The resulting multiplet energies are essentially
the same (see ref. \cite{deb}) as those of Hartree-Fock theory.
\end{thebibliography}
\end{document}